\documentclass[twocolumn,english,prl,amssymb,aps,superscriptaddress,showpacs,twocolumn,amsmath,showkeys,floatfix]{revtex4-1}
\usepackage[T1]{fontenc}
\usepackage[latin9]{inputenc}
\usepackage{geometry}
\usepackage[active]{srcltx}
\usepackage{textcomp}
\usepackage{amsmath}
\usepackage{graphicx}
\usepackage{color}
\usepackage{esint}
\usepackage{svg}
\makeatletter
\usepackage{babel}

\makeatother

\begin{document}
	
	\title {Dynamics of mode-locked nanolasers based on Hermite-Gaussian modes}
	\author{Yifan Sun}
	\affiliation{Universit\'e Paris-Saclay, CNRS, ENS Paris-Saclay, CentraleSup\'elec, LuMIn, 91190 Gif-sur-Yvette, France}
	\author{Sylvain Combri\'e}
	\affiliation {Thales Research and Technology, 91120 Palaiseau, France}
	\author{Alfredo De Rossi}
	\affiliation {Thales Research and Technology, 91120 Palaiseau, France}
	\author{Fabien Bretenaker}
	\affiliation{Universit\'e Paris-Saclay, CNRS, ENS Paris-Saclay, CentraleSup\'elec, LuMIn, 91190 Gif-sur-Yvette, France}
	\affiliation{Light and Matter Physics Group, Raman Research Institute, Bangalore 560080, India}

	\begin{abstract} 
		The different dynamical behaviors of the Hermite-Gaussian (HG) modes of mode-locked nanolasers based on a harmonic photonic cavity are  investigated in detail  using a model based on a modified Gross-Pitaevskii Equation. 
		Such nanolasers are shown to exhibit mode-locking with a repetition rate independent of the cavity length, which is a strong asset for compactness.
		The differences with respect to conventional lasers are shown to originate from the peculiar gain competition between HG modes, which is investigated in details. In the presence of a saturable absorber, the different regimes, i. e. Q-switching, Q-switched mode-locking, and continuous-wave (cw) mode locking, are isolated in a phase diagram and separately described.
		Mode-locking is found to be robust against phase-intensity coupling and to be achievable in a scheme with  spatially separated gain and absorber.
	\end{abstract}
	
	
	\maketitle
	
	\section{Introduction}
	
	The integration of  ultra-compact laser sources on a silicon platform should strongly improve the energy efficiency in short-distance and on-chip communication for future computers by suppressing a large volume of cables and components \cite{Miller2009,Sun2015,Ning2019}. In this context, passively mode-locked semiconductor lasers are a promising source of optical pulses at high repetition rates for integrated photonics, which are potentially compact, low-cost, and reliable. Besides, photonic crystal (PC) lasers, which exhibit appealing properties in terms of low volume, low threshold, and excellent energy efficiency \cite{Park2004,Matsuo2010,Crosnier2017}, have achieved significant progress since the first photonic crystal laser was demonstrated \cite{Painter1999}. By modifying the geometry of the PC, different cavity properties can be optimized, leading for example to ultrahigh quality factors \cite{Song2005} or high collection efficiency \cite{Tran2009}. Moreover, nanolasers based on photonic crystal (PC) cavities are of great importance from the point of view of quantum optics due to their ability to tightly confine modes in three dimensions \cite{Ning2019}. 
	
	With the objective of generating short pulses of light, self-pulsing operation of Fano photonic crystal lasers has been demonstrated \cite{Rasmussen2017b,Yu2017a}. Moreover,  mode-locked operation has been considered for lasers based on  PC cavities \cite{Liu2005,Heuck2010}. Recently, a new type of photonic crystal cavity forming  a harmonic photonic potential leading to the possible oscillation of  multiple longitudinal Hermite-Gaussian (HG) modes has been experimentally demonstrated \cite{Marty2019,Combrie2017a,Poulton2015a}. Based on such harmonic potential cavity, a new concept of ultra-compact mode-locked nanolaser has been proposed \cite{Sun2019}. It is based on the fact that in the presence of a harmonic potential cavity HG modes exhibit a periodic spectrum, which is a necessary condition for mode-locking. 
	Moreover, the inhomogeneous intensity distribution of HG modes inside the cavity, compared to the standing waves of conventional Fabry-Perot cavities, and the compactness of the cavity, lead to
	different scalings for the  laser parameters
	compared to conventional  mode-locked lasers. 
	In particular, the repetition rate of the laser pulse train is governed by the curvature of the photonic potential and not the the cavity length \cite{Sun2019}. Beyond the initial prediction that such harmonic cavity lasers should exhibit  passive mode locking, the aim of the present paper is to give a detailed account of the possible dynamical behaviors predicted for such lasers.
	
	Due to the spatial inhomogeneity of the intensity distribution of HG modes, investigating the dynamical behavior  of such parabolic cavity lasers cannot be easily performed using conventional methods usually used to model mode-locked lasers. For example,  the Haus master equation \cite{Haus2000}, which is widely used in theory of semiconductor mode-locked laser, describes the evolution of the pulse after one round-trip inside the cavity in a reference frame moving with the pulse, and treats the gain and absorber as lumped elements inside the cavity. 	
	 We choose here not to follow this approach because i) it is well adapted to the situation where the pulse duration is much shorter than the cavity round-trip time, which will not be the case in nanolasers that will sustain oscillation of only a few modes, and ii) Haus' master equation is written in a reference frame moving at the pulse group velocity, while we favor a static reference frame.
	Another widely used method is based on delay differential equations \cite{Vladimirov2004}, which easily provides the bifurcation analysis and does not assume small gain and loss per cavity round trip. However, this method is also  not well adapted to nanolasers in which the pulse fills more or less the resonator. We thus describe in Section \ref{Model} below a model based on the Gross-Pitaevskii Equation (GPE) with dissipative terms.
	The cavity effect is based on the second order dispersion and the spatially dependent parabolic potential. The dissipative terms describe the saturable gain and absorption in the active structure. In this framework, the spatial effects typical of HG modes, linked to the spatial distributions of the gain and the absorber inside the resonator, are properly taken into account. 
	
	
	After having derived the model in Section \ref{Model},  
	Section \ref{Competition} is devoted to the investigation of the specific aspects of the competition between the HG modes in the presence of gain saturation. In particular, we isolate the peculiarities of the competition between HG modes compared to the usual modes of a FP cavity and the role of the response time of the active medium in the result of mode competition. Section \ref{Regimes} gives a detailed  description of the different dynamical regimes achievable in the  harmonic cavity nanolaser, namely Q-switched operation, Q-switched mode locking, and cw mode locking. Finally, in Section \ref{Stability}, the influence of different parameters on the stability of the mode-locked soliton oscillation regime are investigated, such as the presence of a non-zero Henry factor in the semiconductor gain and absorber sections. We also investigate whether mode-locking can be obtained using spatially separated gain and absorber sections in the cavity, which could lead to more practical implementations.  Finally, a general comparison between harmonic cavity lasers and  traditional FP cavity lasers is also presented. 
	
	\section{Model}\label{Model}
	\subsection{From the wave equation to the Schr\"odinger equation}
	Let us consider the wave equation in a uni-dimensional nonlinear medium with periodic relative permittivity $\varepsilon_r(x)$:
	\begin{equation}
	\left[\frac{\partial^2}{\partial x^2}-\frac{\varepsilon_r(x)}{c^2}\frac{\partial^2}{\partial t^2}\right]E(x,t)=\mu_0\frac{\partial^2P_{\rm{NL}}}{\partial t^2}\ ,
	\label{eq_01}
	\end{equation}
	where $c$ is the light velocity in vacuum and $P_{\rm{NL}}$ is the nonlinear polarization. Sipe and Winful have theoretically demonstrated, using the multiple scales method and Floquet-Bloch theory \cite{Sipe1988,DeSterke1988}, that, in the case of a Kerr nonlinearity, eq.\,\eqref{eq_01} can be replaced by the nonlinear Schr\"odinger equation (NLSE): 
	\begin{equation}
	i\frac{\partial A}{\partial t} +\frac{1}{2}\omega_{kk} \frac{\partial^2A}{\partial x^2} + \alpha |A|^2A=0\ .
	\label{eq_02}
	\end{equation}
	In this equation, $A(x,t)$ is the slowly varying amplitude of the field related to $E(x,t)$ through
	\begin{equation}
	E (x,t)=A(x,t)u(x)e^{i k x}e^{-i \omega_0 t}+\mathrm{c.c.}\ ,\label{eq_03}
	\end{equation}
	where  $u(x)$ is periodic with the same period as $\varepsilon_r(x)$, $\omega_{kk}=\partial^2\omega/\partial k^2$ is the group velocity dispersion, $\alpha$ is the effective non linearity seen by the field envelope, and $\omega_0$ is the center frequency. 
	The group velocity dispersion can be largely controlled by the PC structure \cite{Notomi2001}. The band edge is located at a high-symmetry point in the reciprocal space in most cases. Therefore, within the spectral domain of interest, high order dispersion is controllable \cite{Sun2019}.
	
	\subsection{Harmonic photonic cavity description: Gross-Pitaevskii Equation (GPE)}
	One can create an effective potential for light by spatially varying one parameter of a dielectric guiding nanostructure, for example the period $a$ of the confining holes, along a given direction $x$. This formalism holds in the limit of slow changes of $a$, as shown mainly by experimental and numerical verification \cite{Combrie2017a,Marty2019} but also by some theoretical arguments \cite{Vigneron2007,Dobbelaar2015}. A minimum in the effective photonic potential can thus be used as a resonator to confine light, whose evolution is then governed by the Gross-Pitaevskii Equation (GPE), which is constructed by adding the potential $V(x)$ to the NLSE:
	\begin{equation}
	\dot{\imath}\frac{\partial A}{\partial t} +\frac{1}{2}\omega_{kk} \frac{\partial^2A}{\partial x^2} -V(x) A = \alpha |A|^2A\,.\label{eq_04}
	\end{equation}
	One example of such a resonator design is the one of a chirped periodic dielectric material with a relative permittivity $\varepsilon_r(x)=\bar{\varepsilon}+\Delta\varepsilon\cos(2\pi x/a(x))$ in which the period $a(x)$ slowly changes along $x$ according to a parabolic evolution, namely $a(x)=a_0+\varsigma x^2$. The limitation to small change of $a$ leaves the normal modes approximately unchanged, as well as $\omega_{kk}$. Equation (\ref{eq_04}) still holds, but the spatial change of $a$ induces a frequency offset $V(x)\propto [a(x)^{-1} - a_0^{-1}]\propto -\varsigma x^2$. Hence, such a chirped periodic dielectric results into a harmonic photonic potential for the field envelope of the normal modes near the band edge. This idea was proposed in a slightly different implementation involving the modulation of the thickness of a patterned slab \cite{Dobbelaar2015}.
	
	
	Applying this to the GPE eq. \eqref{eq_04} to the ``cold cavity'' without any Kerr nonlinearity ($\alpha=0$) indeed leads to the Schr\"odinger equation for the 1D quantum harmonic oscillator:
	\begin{equation}
	i\frac{\partial A}{\partial t} +\frac{1}{2}\omega_{kk} \frac{\partial^2A}{\partial x^2} -\frac{1}{2}\frac{\Omega^2}{\omega_{kk}}x^2 A =0\ ,
	\label{eq_05}
	\end{equation}
	where $\Omega$ is the free spectral range of the modes and $\Omega^2/\omega_{kk}^2$ characterizes the curvature of the harmonic potential. 
	The eigensolutions are the Hermite-Gaussian (HG) functions 
	\begin{equation}
	\Psi_n(x)=\frac{1}{\sqrt{2^n n!}}\pi^{-1/4}\exp (-x^2/2)H_n(x)\ ,
	\label{eq_06}
	\end{equation}
	with
	\begin{equation}
	H_n(x) = n!\sum_{m=0}^{\lfloor\frac{n}{2}\rfloor} \frac{(-1)^m}{m!(n-2m)!}(2x)^{(n-2m)}\ ,
	\label{eq_07}
	\end{equation}
	where $\lfloor\frac{n}{2}\rfloor$ denotes the largest integer less than or equal to $\frac{n}{2}$.
	%
	Such modes can exhibit high-Q values, as experimentally demonstrated \cite{Poulton2015,Combrie2017a,Marty2019}. These localized states are the same as the solutions of the quantum harmonic oscillator, as shown in Fig.\,\ref{fig:mode_wide_and_gain_width}(a). Their electromagnetic energy distributions are very different from those of the homogeneous standing wave modes of standard Fabry-Perot cavities. Moreover, the frequency separation between the HG modes depends on the effective mass $m^{-1}_{\rm eff}=\hbar^{-1}\partial^2_k\omega=\hbar^{-1}\omega_{kk}$ of the quasi-particle and is related to the potential stiffness $V(x)=\frac{1}{2}m_{\mathrm{eff}}\Omega^2x^2/\hbar=\frac{\Omega^2}{2\omega_{kk}}x^2$, instead of the size of the oscillator.  
	This can be an effective way to reduce the cavity size for a given desired value of the mode frequency separation $\Omega$. The size of the cavity can be determined from the scaling factor $x_\Omega=\sqrt{\omega_{kk}/\Omega}$, which is the width of the fundamental HG mode. Setting for example $\Omega/2\pi=100\,\rm{GHz}$ and taking a typical value for the dispersion $\omega_{kk}=2 v_g^2/\Delta\omega_g=45\,\rm{m^2\cdot rad\cdot s^{-1}}$ estimated from the typical photonic bandgap $\Delta\omega/\omega\approx20\%$ of the PC cavity and the group velocity $c_0/4$ in semiconductor waveguides, leads to a size $x_\Omega=8.4\,\mu\mathrm{m}$. 
	
	The field $A(x,t)$ inside the cavity can be expanded on the basis constituted by these Hermite-Gaussian modes
	\begin{equation}
	A(x,t)=\sum_{n=0}^{\infty}C_n(t)e^{-i\omega_n t}\Psi_n(x)\ ,
	\label{eq_field_expension}
	\end{equation}
	where $\omega_n=(n+1/2)\Omega$. For a given field distribution $A(x,t)$, the modal coefficient $C_n(t)$ can be calculated by projecting it on $\Psi_n$: 
	\begin{equation}
	C_n(t)=e^{i\omega_n t}\int_{-\infty}^{\infty}A(x,t)\Psi_n(x)dx.
	\label{eq_09}
	\end{equation}

	\subsection{Harmonic cavity laser: dissipative terms}
	The cold cavity described by Eq.\,\eqref{eq_05} can be transformed into a laser by adding gain inside or hybridized to the cavity. Moreover, mode-locking can be favored by adding a saturable absorber. Adding in an empirical manner these elements to Eq.\,\eqref{eq_05} leads to the so-called modified GPE:
	\begin{equation}
	i\frac{\partial A}{\partial t} +\frac{1}{2}\omega_{kk} \frac{\partial^2A}{\partial x^2} -\frac{1}{2}\frac{\Omega^2}{\omega_{kk}}x^2 A -iH_1(|A|^2)A=0\ .
	\label{eq_master_equation}
	\end{equation}
	The dissipative term $H_1$ describes the gain and absorption according to
	\begin{equation}
	H_1=\frac{1}{2}g(x,t)(1-\mathrm{i}\alpha_{\mathrm{g}}) - \frac{1}{2}a(x,t)(1-\mathrm{i}\alpha_{\mathrm{a}})-\frac{1}{2}\gamma_0\ ,\label{eq_dissipative_H_1}
	\end{equation}
	where $g(x,t)$ and $a(x,t)$ are the time and space dependent gain and saturable absorber coefficients, with their respective Henry factors $\alpha_{\mathrm{g}}$ and $\alpha_{\mathrm{a}}$.  The term $\gamma_0$ holds for the intrinsic losses.
	
	Incidentally, we notice that Eq.~(\ref{eq_master_equation}) is similar to Haus' master equation in the limit case of zero group velocity. But the assumptions of the two models are different.
	
	The fact to use the standard form of Eq.\,(\ref{eq_dissipative_H_1}) for the gain and absorber term is based on the usual approximation that these effects are a small perturbation to the laser, which do not modify the shape of the modes. The modes are supposed to be defined by the ``cold'' resonator only.
	
	
	In general, saturation of the gain and of the absorption is described by the following set of spatially local equations 
	\cite{Agrawal1989}:
	\begin{equation}
	\frac{\partial g(x,t)}{\partial t} = -\frac{g(x,t)-g_0(x)}{\tau_{\mathrm{g}}} - \frac{|A(x,t)|^2}{\tau_{\mathrm{g}}I_{\mathrm{sat,g}}} g(x,t)\ ,
	\label{eq_gain_noninstantaneous}
	\end{equation}
	\begin{equation}
	\frac{\partial a(x,t)}{\partial t} = -\frac{a(x,t)-a_0(x)}{\tau_{\mathrm{a}}} - \frac{|A(x,t)|^2}{\tau_{\mathrm{a}}I_{\mathrm{sat,a}}} a(x,t)\ ,
	\label{eq_absorber_noninstantaneous}
	\end{equation}
	where $\tau_{\mathrm{g}}$ and $\tau_{\mathrm{a}}$ are the lifetimes of the gain and the absorption, respectively, $I_{\mathrm{sat,g}}$ and $I_{\mathrm{sat,a}}$ their saturation intensities, and $g_0$ and $a_0$ the unsaturated values of $g$ and $a$. In the case where the lifetimes $\tau_{\mathrm{g}}$ and/or $\tau_{\mathrm{a}}$ are much shorter than all the response times of the system, $g$ and/or $a$ can be considered to reach steady-state instantaneously, leading to: 
\begin{equation}
	g(x,t)=g_0(x)/\left(1+\frac{{|A(x,t)|^2}}{I_{\rm{sat,g}}}\right),
	\label{eq_14}
	\end{equation}
	\begin{equation}
	a(x,t)=a_0(x)/\left(1+\frac{{|A(x,t)|^2}}{I_{\rm{sat,a}}}\right).
	\label{eq_15}
\end{equation}

    In the following the simulations are performed using a split-step Fourier method with adaptive step size control.  The spatial discretization period is equal to $0.13x_\Omega$. Time discretization is variable but has at least 100 samples per period $2\pi/\Omega$.
    
The initial conditions that we use are a small fraction of $g_0(x)$ for $g(x,0)$ and random complex numbers corresponding to a small intensity for the fields $A(x,0)$.
    
	

	%
	%
	
	\section{Gain saturation: Hermite-Gaussian modes competition}\label{Competition}
\begin{figure}[htb]
		\centering %
		\includegraphics[width=\columnwidth]{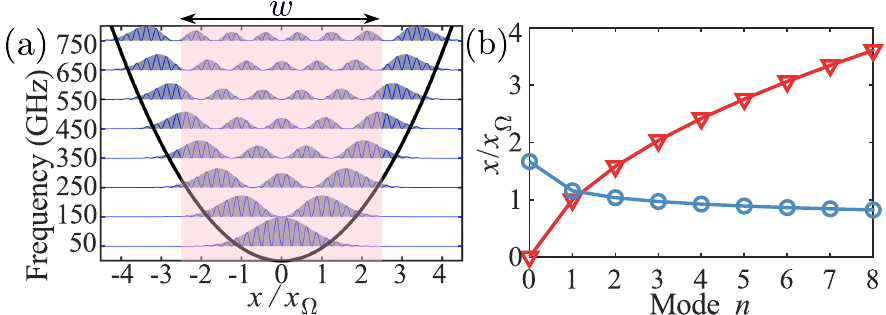}
		\caption{(a)  Shape of the Hermite-Gaussian modes with parabolic potential. The semi-transparent pink area is the active medium, gain or absorber.(b) Location (triangles) and width (circles) of the outermost lobe of HG modes as a function of the mode order $n$. 	}
		\label{fig:mode_wide_and_gain_width}
	\end{figure}

	The peculiarity of the spatial distribution of the light intensity in the case of HG modes, as can be seen in Fig.\,\ref{fig:mode_wide_and_gain_width}, allows us to expect a competition behavior different from the the case of usual Fabry-Perot or ring cavities. The aim of this section is to investigate this competition for gain among HG modes in the absence of saturable absorption.
	\subsection{Mode saturation matrix}
\begin{figure}[htb]
		\centering %
		\includegraphics[width=1\columnwidth]{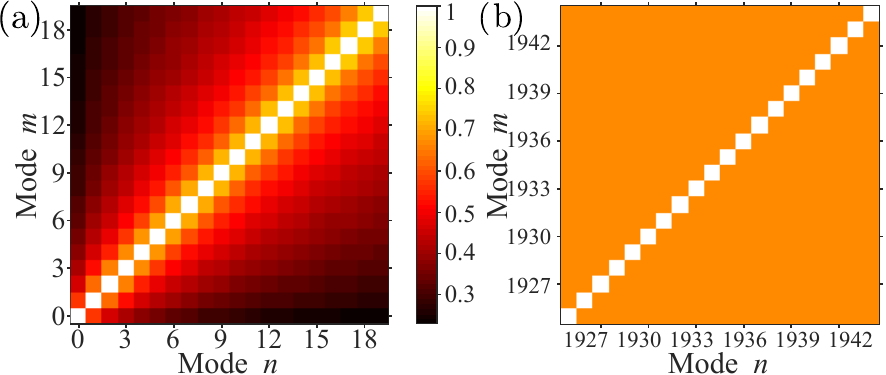}
		\caption{Saturation Matrix $S_{n,m}$ for (a) Hermite-Gaussian modes in harmonic photonic laser and for (b) sinusoidal modes Fabry-Perot laser. The saturation matrix in (a) will remain valid in the case of Section \ref{Asymmetric} where the gain medium fills only one half of the cavity.}
		\label{fig:saturation_matrix}
	\end{figure}

	The spatial inhomogeneity of the intensity distribution of HG modes is very different from the spatial homogeneous distribution of standing waves in FP lasers, so that the cross saturation of the gain of one mode by another mode should be different from what happens in FP lasers. To evaluate this cross-gain saturation, we define the saturation matrix as 
	\begin{equation}
	S_{n,m}=\frac{\int_{-\infty}^{\infty}\Psi_n^2(x)\Psi_m^2(x)dx}{\sqrt{\int_{-\infty}^{\infty}\Psi_n^4(x)dx\int_{-\infty}^{\infty}\Psi_m^4(x)dx}}\ ,
	\label{eq_saturation_matrix_Snm}
	\end{equation}
	where $\Psi_n(x)$ is the spatial dependence of the field envelope of mode of order $n$, i. e. a Hermite-Gaussian mode like in Eq.\,(\ref{eq_06}) for the harmonic photonic cavity or a standing wave for the Fabry-Perot cavity. 
	In both cases the modes are normalized according to:
	\begin{equation}
	\int_{-\infty}^{\infty}\Psi_n(x)\Psi_m(x)dx = \delta_{nm}.
	\label{eq_orthogonality_relation}
	\end{equation}
	The definition of Eq.\,(\ref{eq_saturation_matrix_Snm}) supposes that the gain medium homogeneously fills the resonator. 
	Figure \ref{fig:saturation_matrix} compares the value of the  coefficients of the saturation matrix for the two types of modes. The calculation for standing waves is based on  10 successive modes of a FP semiconductor laser with $100\,\rm GHz$ free spectral range (FSR), operating at $1.55\,\rm \mu m$ wavelength, and filled with a medium of  refractive index equal to $3.5$. 
	
	In both cases, the values of the elements of the diagonal are equal to 1. In the case of the HG modes (see Fig.\,\ref{fig:saturation_matrix}(a)), the cross-saturation coefficients progressively decrease with the distance from the diagonal. This is consistent with the plot of Fig.\,\ref{fig:mode_wide_and_gain_width}(b), which shows that the position of the main lobe of HG mode of order $n$ increases roughly like $\sqrt{n}$. On the contrary, for the FP cavity modes (see Fig.\,\ref{fig:saturation_matrix}(a)), all non diagonal elements are equal to  $0.667$, which means that the cross saturation is the same for all pairs of modes in the FP cavities.
	
	The plots of Fig.\,\ref{fig:saturation_matrix} permit us to predict that mode competition among HG modes will be different from the one experienced in  usual FP lasers. Indeed, while in the latter case one dominant mode will equally saturate the gain seen by all other modes, competition among HG modes will be fierce only among neighbouring modes. Therefore, one can expect mode of very different orders to be quite easily able to oscillate simultaneously. 
	\subsection{Influence of the gain window width}
	In the laser geometry that we consider, which is sketched in Fig.\,\ref{fig:mode_wide_and_gain_width}(a), the gain window has a width $w$. Since the width of HG mode of order $n$ scales roughly like $\sqrt{n}$, we can control the number of modes that compete for the gain by changing $w$. 
	
	
	\begin{figure}[htb]
		\centering
		\includegraphics[width=\columnwidth]{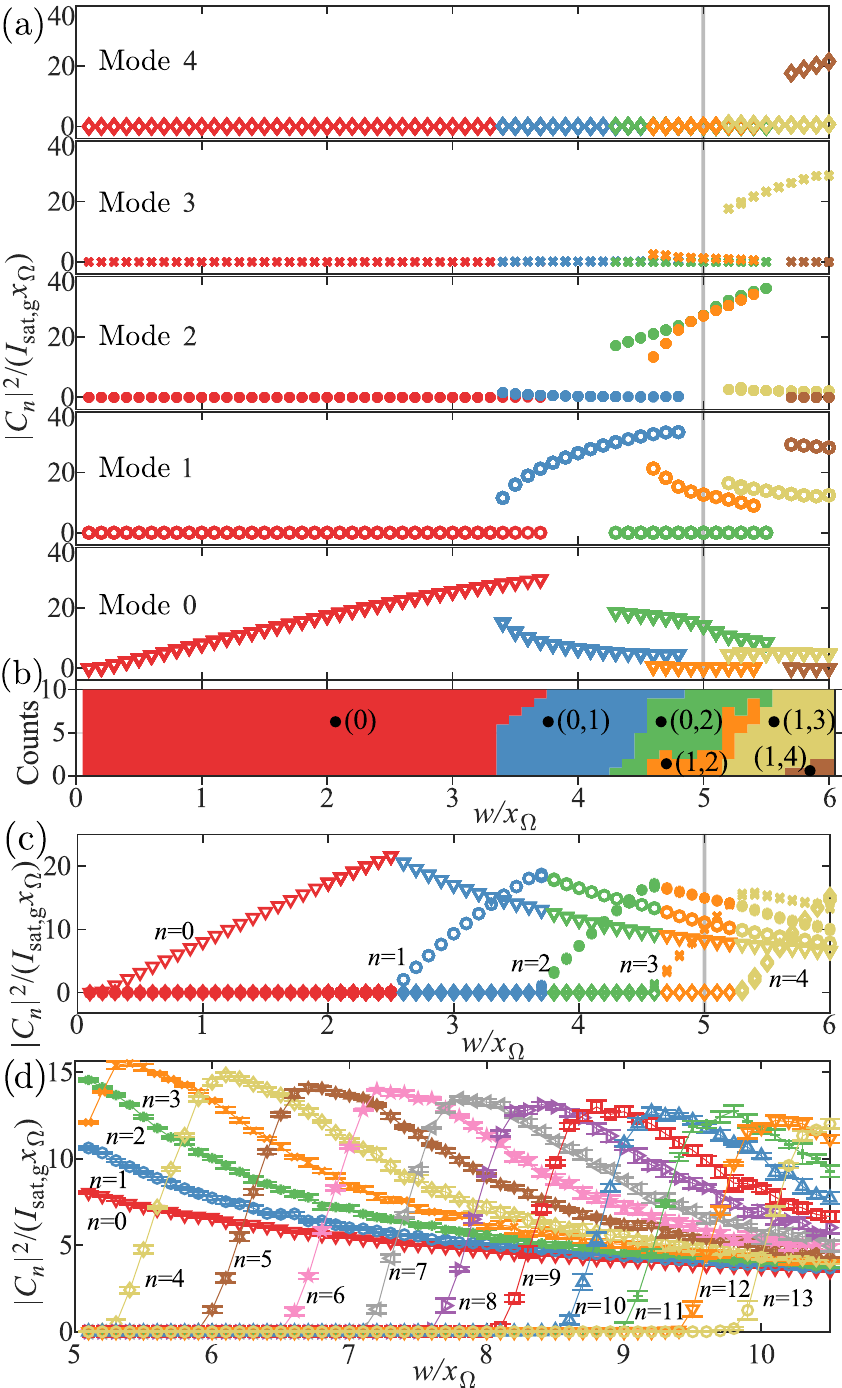}	
		\caption{Evolution of the steady-state mode intensities $|C_n|^2$ as a function of the gain window width $w$. The simulation is run 10 times with random initial fields for each case. (a) Instantaneous gain saturation. Evolution of mode intensities versus $w$ for $n=0..4$. (b) Corresponding counts for each solution versus $w$. The numbers in parentheses represent the orders of oscillating HG modes. Each color in (a) and (b) represents one possible steady-state. (c) Gain with finite lifetime $\tau_\mathrm{g}=1\,\mathrm{ns}$. Evolution of mode intensities versus $w$ for $n=0..4$. Simulation is run 10 times in each case. (d) Same as (c) for larger values of $w$. Each color in (d) represents one mode. There is no multistability in this case. Parameter values: $g_0=10\,\gamma_0$, $a_0=0$. } 
		\label{fig:mode_intensity_counts_vs_gain_width}
	\end{figure}
	
	We thus suppose that the gain region exhibits a homogeneous unsaturated gain coefficient coefficient $g_0=10\,\gamma_0$, with $\gamma_0=10^{10}\,\mathrm{s}^{-1}$, centered on the bottom of the photonic potential with a width $w$,  as shown by the semi-transparent region in Fig. \ref{fig:mode_wide_and_gain_width}(a). To avoid boundary effects, we replace the walls of the rectangular gain window by a smoothed function. We also suppose that there is no saturable absorber inside the cavity, i. e. $a_0=0$. The steady-state laser fields are obtained by running the calculations with a fixed gain length $w$ and starting from random initial field amplitudes with a maximum intensity equal to $0.0001I_{\rm sat,g}$. The simulation is run until the field amplitudes reach their steady-state values. The steady-state mode intensities $|C_n|^2$ are then obtained by expending the field $A(x,t)$ using Eq. (\ref{eq_09}). 
	
	We start by considering the case of instantaneous gain saturation, as shown in Eq. (\ref{eq_14}). For each value of $w$ ranging from $0$ to $6\,x_\Omega$, the calculation is run 10 times. 
	
	Figure \ref{fig:mode_intensity_counts_vs_gain_width}(a) reproduces the evolution of the steady-state intensities of the first 5 modes as a function of $w$. The number of occurrences of the different steady-state solutions are summarized by color regions in Fig. \ref{fig:mode_intensity_counts_vs_gain_width}(b), in which the numbers marked in brackets represent the orders of the dominant modes for each region.
	
	We can see that in the region of $0<w/x_\Omega<3.3$, only 
mode 0 oscillates, because  the gain region is to narrow to sustain oscillation of higher order modes. Then, for $3.4<w/x_\Omega<3.8$, two steady-state solutions prevail: one for which mode 0 oscillates alone, and one in which mode 0 and mode 1 oscillate simultaneously. Further, when $w/x_\Omega\geq 3.8$, the solution for which mode 0 is alone is no longer stable, and simultaneous oscillation of modes 0 and 1 is stable till $w/x_\Omega\geq 4.8$. Then, for $w/x_\Omega\geq 4.4$, simultaneous oscillation of modes 0 and 2 occurs and  disappears for $w/x_\Omega>5.5$. The simultaneous oscillation of  mode 1 and mode 2 can occur in the region of $4.6\leq w/x_\Omega\leq 5.4$. 

We can thus see that, in the case of instantaneous gain saturation,  the competition among  HG modes leads to a complicated multi-stability situation. The general tendency is that higher order modes are favored when $w$ is increased, but the number of stable steady-state solutions also increases with $w$. 
	
The situation becomes completely different when we introduce the non-instantaneous response of the gain described by Eq.\,(\ref{eq_gain_noninstantaneous}). For example,  Fig.\,\ref{fig:mode_intensity_counts_vs_gain_width}(c) reproduces the results obtained with a gain lifetime equal to $\tau_\mathrm{g}=\rm{1ns}$. All other parameter values are kept equal to the case of instantaneous saturation of Figs.\,\ref{fig:mode_intensity_counts_vs_gain_width}(a,b) and  the simulation is also run 10 times starting from random initial conditions for each situation. We can see that in this situation the multi-stability of Fig.\,\ref{fig:mode_intensity_counts_vs_gain_width}(b) disappears: only one steady-state solution is obtained for each value of $w$. 	For $w/x_\Omega\leq 2.6$, only mode 0 oscillates, just like in the case of instantaneous gain saturation. But it is clear that, for larger values of $w$, a larger number of modes can simultaneously oscillate than for instantaneous gain saturation. For example, for $w=6x_{\Omega}$, all the five modes considered here oscillate simultaneously. This tendency is confirmed in Fig.\,\ref{fig:mode_intensity_counts_vs_gain_width}(d) for $w$ up to $10\,x_{\Omega}$, where the 14 first modes can oscillate simultaneously. The effect of finite gain lifetime is thus clearly to reduce competition between the HG modes. We can expect this effect to be favorable to stable mode-locking operation in the presence of a saturable absorber.

\subsection{Influence of gain lifetime}

\begin{figure}[htb]
		\centering 
		\includegraphics[width=\columnwidth]{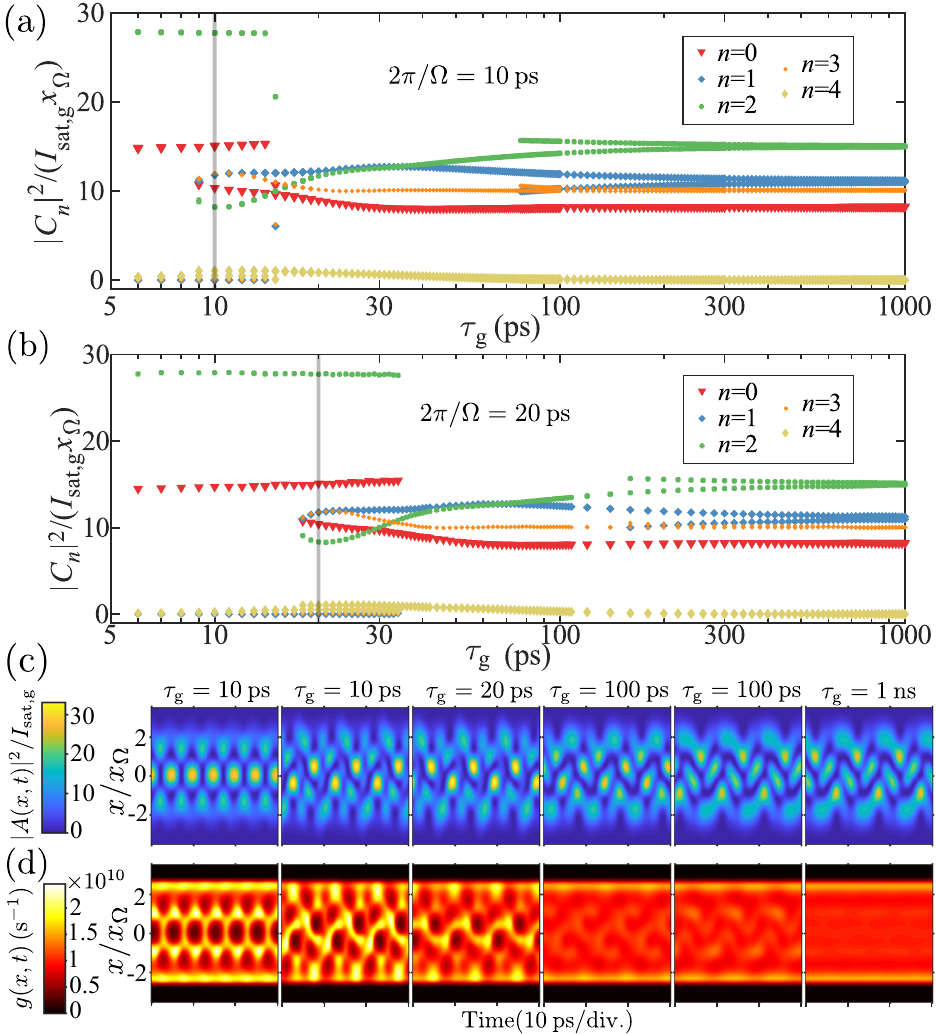}		
		\caption{(a,b) Steady-state mode intensities as a function of  gain lifetime $\tau_\mathrm{g}$. The simulation is run 10 times with random initial conditions and with $w=5\,x_\Omega$ and $g_0=10\,\gamma_0$, $a_0=0$. Mode separation is (a) $\Omega/2\pi=100\,\mathrm{GHz}$ and (b) $\Omega/2\pi=50\,\mathrm{GHz}$. (c,d):  False color plots of (c) the field intensity distribution $|A(x,t)|^2$ and (d) the gain distribution $g(x,t)$ versus time (horizontal axis) and position $x$ (vertical axis) in steady-state for different values of $\tau_{\mathrm{g}}$. The values of the other parameters are the same as in (a).}
		\label{fig:mode_intensity_vs_gain_lifetime}
\end{figure}	

Based on the difference between the results obtained for instantaneous and non-instantaneous saturation of the gain in Fig.\,\ref{fig:mode_intensity_counts_vs_gain_width}, we can infer that the gain lifetime has a strong influence on the laser steady-state behavior. To further investigate this dependence, we choose a fixed gain width $w=5\,x_\Omega$ and run the simulation 10 times starting from random initial fields with maximum intensity $0.0001I_{\rm{sat,g}}$ and  excitation ratio equal to $r_\mathrm{e}=g_0/\gamma_0=10$ for $\tau_\mathrm{g}$ ranging from $6\,\rm{ps}$ to $1000\,\rm{ps}$. 
	
The corresponding steady-state mode intensities are reproduced in Figs.\,\ref{fig:mode_intensity_vs_gain_lifetime}(a) and \ref{fig:mode_intensity_vs_gain_lifetime}(b) for  mode separation values of $\Omega/2\pi=100\,\rm{GHz}$  and $\Omega/2\pi=50\,\rm{GHz}$, respectively. These results clearly show the transition between the two regimes, i. e., instantaneous gain saturation when $\tau_\mathrm{g}\ll 2\pi/\Omega$ and slow gain saturation when $\tau_\mathrm{g}\gg 2\pi/\Omega$. For example, in Fig.\,\ref{fig:mode_intensity_vs_gain_lifetime}(a), the transition between the two regimes occurs for  $8\,\rm{ps}\lesssim\tau_\mathrm{g}\lesssim 17\,\rm{ps}$, which is  consistent with the value $2\pi/\Omega=10\,\rm{ps}$. This is confirmed by Fig.\,\ref{fig:mode_intensity_vs_gain_lifetime}(b), for which we have taken $2\pi/\Omega=20\,\rm{ps}$. Then the transition region is shifted accordingly to $17\,\mathrm{ps}\lesssim\tau_\mathrm{g}\lesssim 33\,\mathrm{ps}$. The mode separation $\Omega$, which is the frequency of the beatnote between the modes, is thus a key factor to understand the influence of gain dynamics on the laser competition behavior. If this frequency is much larger than $1/\tau_\mathrm{g}$, the gain saturation can no longer follow the beatnote between the modes, and conversely. 

The transition between the two regimes can also be directly observed by looking at the spatiotemporal distribution of the gain inside the laser. This is done in Figs.\,\ref{fig:mode_intensity_vs_gain_lifetime}(c,d), which reproduce the evolution versus $x$ and $t$ of the intensity and the gain inside the laser for different values of $\tau_{\mathrm{g}}$ in the case $2\pi/\Omega = 10\,\mathrm{ps}$. One can clearly see that the pattern created by the beatnote between the modes (see Fig,\ref{fig:mode_intensity_vs_gain_lifetime}(c)) is imprinted in the gain distribution only as long as $\tau_{\mathrm{g}}$ is shorter or of the order of $2\pi/\Omega$. On the contrary, when $\tau_{\mathrm{g}}$ is much longer than $2\pi/\Omega$, the gain is saturated more uniformly, and the spatiotemporal hole burning disappears.

In addition, we verified that the linear loss rate $\gamma_0$ is not a key parameter for this transition. Indeed, we checked that when we change $\gamma_0$ from $10^{10}\,\mathrm{s}^{-1}$ to $15\times10^{10}\,\mathrm{s}^{-1}$, the results of  Fig.\,\ref{fig:mode_intensity_vs_gain_lifetime}(a) are almost unchanged. In particular, the transition between the two regimes always occurs when $\tau_{\mathrm{g}}$ is close to $2\pi/\Omega$.

Some multistability occurs at the transition region, i. e., when the repetition period $2\pi/\Omega$ is close to the gain lifetime. Another multistability zone can be seen for $\tau_{\mathrm{g}} \simeq 75\,\rm{ps}$ in Fig.\,\ref{fig:mode_intensity_vs_gain_lifetime}(a) and $160\,\rm{ps}$ in Fig.\,\ref{fig:mode_intensity_vs_gain_lifetime}(b). These multiple states converge to the same one when the gain lifetime is increased.

\section{Detailed description of the different oscillation regimes}\label{Regimes}
\begin{figure}[htb]
		\centering
		\includegraphics[width=1\columnwidth]{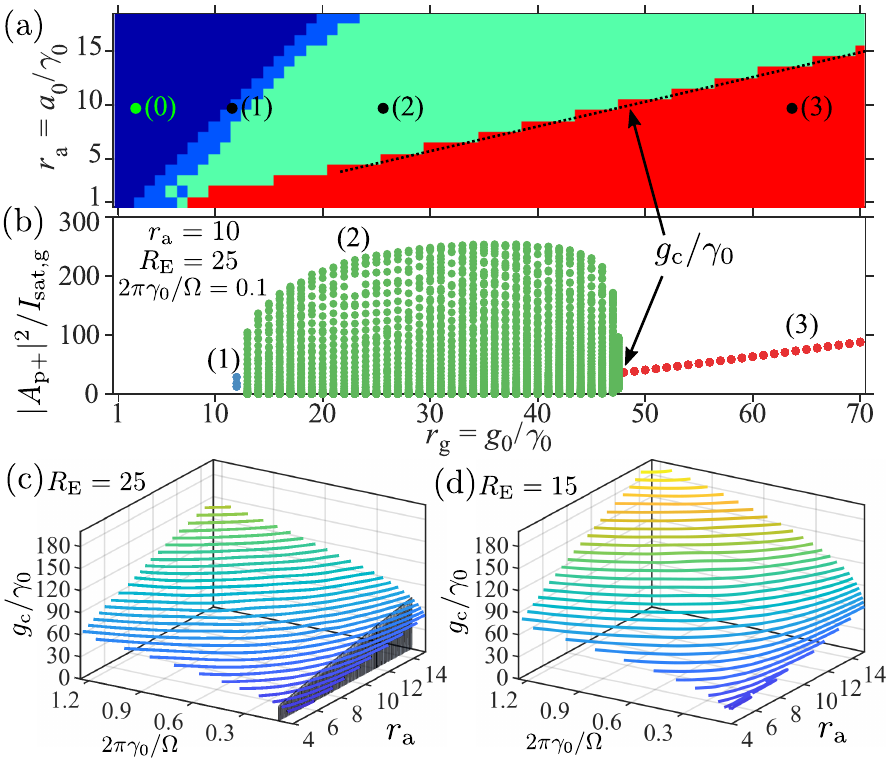}	
		\caption{(a) Phase diagram representing the different steady-state regimes as a function of  unsaturated gain and absorption normalized to $\gamma_0$:  (0) below threshold; (1) Q-switched operation; (2) Q-switched ML operation; (3) cw mode locking. (b) Bifurcation diagram:  peak values $|A_{\rm p+}|^2$ of the field intensity propagating in $+x$ direction at $x=0$ in the steady-state as a function of $r_{\rm{g}}$. The absorption is fixed at $r_\mathrm{a}=a_0/\gamma_0=10$.
(c) The contours of gain pumping rate $g_c/\gamma_0$ between the CW mode locking and Q-switched mode locking as a function of the ratio $2\pi\gamma_0/\Omega$ and absorption rate ratio $r_a$ for the saturation energy ratio of $R_\mathrm{E}=25$. (d) The same situation with the saturation energy ratio $R_\mathrm{E}=15$. The common parameter $\rm\Omega/2\pi=100GHz$ is fixed.		
		}\label{fig:bifurcation_peaks}
\end{figure}
The dynamical behavior of the harmonic photonic mode-locked laser in the presence of a saturable absorber has been described in Ref.\,\cite{Sun2019}. In particular, mode-locked pulsed operation has been predicted, which corresponds to the oscillation of a stable dissipative soliton. In the limit where gain saturation is instantaneous, this soliton has been found to perfectly match  the coherent state of a quantum mechanical harmonic oscillator. However, many other  dynamical behaviors have also been predicted \cite{Sun2019}. The aim of the present section is to investigate these regimes in details.
	
We suppose here for the sake of simplicity that the gain and absorber share the same region centered on the potential minimum with a width $w=5\,x_\Omega$, as shown by the semi-transparent region of Fig. \ref{fig:mode_wide_and_gain_width}(a). The lifetime of the gain and the absorber are respectively $1\,\rm{ns}$ and $10\,\rm{ps}$ \cite{Heuck2010,Vladimirov2009b}. The ratio of saturation energies is taken to be $R_E=I_{\rm{sat,g}}\tau_{\rm{g}}/I_{\rm{sat,a}}\tau_{\rm{a}}=25$ \cite{Heuck2010}. In such a non-instantaneous gain and absorber saturation situation, several different dynamical behaviors are observed when the values of $g_0$ and $a_0$ are tuned, as summarized by the phase diagram Fig.\,\ref{fig:bifurcation_peaks} (a).
	
This diagram can be divided into 4 regions: (0) corresponds to no lasing (deep blue), (1) to Q-switching operation (light blue), (2) to Q-switched mode-locked operation (green), and (3) to cw mode locking (red). The borders between these regions look like straight lines in the $\{g_0,a_0\}$ plane. 

The influence of the linear loss rate $\gamma_0$ on the transition from Q-switched mode-locked operation to cw mode locking is investigated in Figs.\,\ref{fig:bifurcation_peaks}(c,d). The critical value $g_c$ of $g_0$ at which it occurs (see Fig.\,\ref{fig:bifurcation_peaks}(a)) increases with $\gamma_0$ and also with $r_{\mathrm{a}}$, as can be seen from Fig.\,\ref{fig:bifurcation_peaks}(c). The vertical cross-section in 
Fig.\,\ref{fig:bifurcation_peaks}(c) for $2\pi\gamma_0/\Omega=0.1$ corresponds to the line separating the green region labeled (2) from the  red region labeled (3) in Fig.\,\ref{fig:bifurcation_peaks}(a). The plot of Fig.\,\ref{fig:bifurcation_peaks}(d) is similar to the one of Fig.\,\ref{fig:bifurcation_peaks}(c) with a smaller value of $R_E$.
	
These four different regions can also be visualized by plotting the peak intensity for the field traveling in the $+x$ direction inside the cavity when the laser is in steady-state regime. This is presented in Fig.\,\ref{fig:bifurcation_peaks}(b) as a function of $g_0$ for a fixed value of $a_0$ equal to $10\,\gamma_0$. In each situation, after steady-state is reached in the simulation, 
the field $A_{+}(x,t)$ propagating in the $+x$ direction is calculated by taking the spatial Fourier transform  $A(k,t)$ of $A(x,t)$ in the $k$-space, then filtering the part of the field field with $k>0$, and transforming it back into $A_+(x,t)$.	The peak values of the field intensity $|A_{\rm p+}|^2$ propagating in the $+x$ direction at the cavity center $x=0$ are then detected within a time duration of $10\,\rm{ns}$ and are plotted in Fig.\, \ref{fig:bifurcation_peaks}(b), which is thus a bifurcation diagram of the laser dynamics. The different regimes are described in detail in the following subsections.
	
\subsection{Q-switched operation}

\begin{figure}[htb]
		\centering
		\includegraphics[width=1\columnwidth]{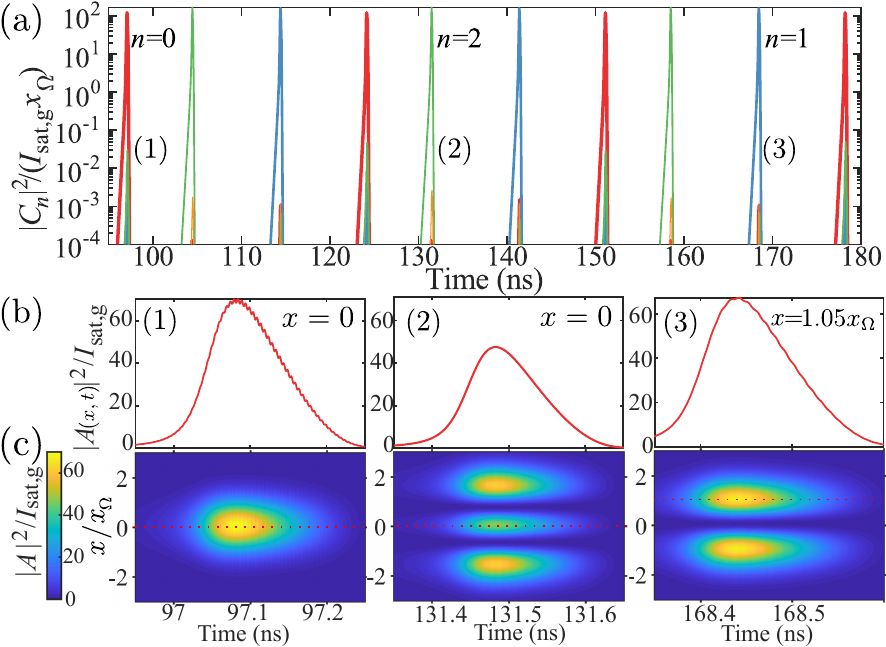}		
		\caption{Q-switched laser behavior. (a) Mode intensities as a function of time. (b) Time evolution of field intensity $|A(x,t)|^2$ at different locations $x$ at three instants labeled 1), (2) and (3) in (a). (c) Corresponding time space map of the intensity $|A(x,t)|^2$. The parameters are  $r_\mathrm{g}=g_0/\gamma_0=12$, $r_\mathrm{a}=a_0/\gamma_0=10$, $w=5\,x_\Omega$, $\tau_g=1\,\mathrm{ns}$, $\tau_a=10\,\mathrm{ps}$.}
		\label{fig:single_mode_q_switching}
	\end{figure}

In the Q-switching region labeled (1) in Fig.\,\ref{fig:bifurcation_peaks}(b), one can see that there are three points. This corresponds to a periodic series of three pulses of three different peak powers, as shown in Fig.\,\ref{fig:single_mode_q_switching}. 

	
	One example of steady-state laser behavior in this region, corresponding to  $r_\mathrm{g}=g_0/\gamma_0=12$, $r_\mathrm{a}=a_0/\gamma_0=10$, and $x/x_\Omega=5$, is shown in details in Fig.\,\ref{fig:single_mode_q_switching}. 
	Q-switching happens here when the resonator losses are kept at a relatively high level compared with gain, allowing the active medium to accumulate a large gain before Q-switching occurs. Once the laser starts, the pulse builds up very quickly and suddenly saturates the absorber down to  small absorption values. 
	
Figure\,\ref{fig:single_mode_q_switching}(a) shows the time evolution of mode intensities $|C_n|^2$ in red steady-state,  calculated using Eq. (\ref{eq_09}) by projecting field $A(x,t)$ on the basis of HG modes. 
One can see the first three modes, corresponding to $n= 0$, 1, and 2, alternately oscillate. The delay between two pulses is about $10\,\rm{ns}$, but it is slightly shorter between modes 0 and 2 than between the other modes. One can also notice that  the intensity of mode 0 is slightly smaller than the other modes, because this mode is spatially smaller than the other ones (see Fig.\,\ref{fig:mode_wide_and_gain_width}) and thus bleaches the gain for smaller peak energies. Thus, after emission of a pulse in mode 0, some gain is left available for mode 2, which can oscillate a bit earlier.

Figure\,\ref{fig:single_mode_q_switching}(b) reproduces the time evolution of the total intensity $|A(x,t)|^2$ at the cavity center $x=0$ for pulses labeled (1) and (2) and at $x=1.05\,x_{\Omega}$ for pulse labeled (3). The detailed evolution of $|A(x,t)|^2$ versus $x$ and $t$ for these three pulses are shown in Fig.\,\ref{fig:single_mode_q_switching}(c). The duration of each  pulse is of the order of $0.1\rm{ns}$. The pulse rise time is shorter than its decay time, as is typically obtained in Q-switched laser. 
	
\subsection{Q-switched mode-locking}
	\begin{figure}[htb]
		\centering
		\includegraphics[width=1\columnwidth]{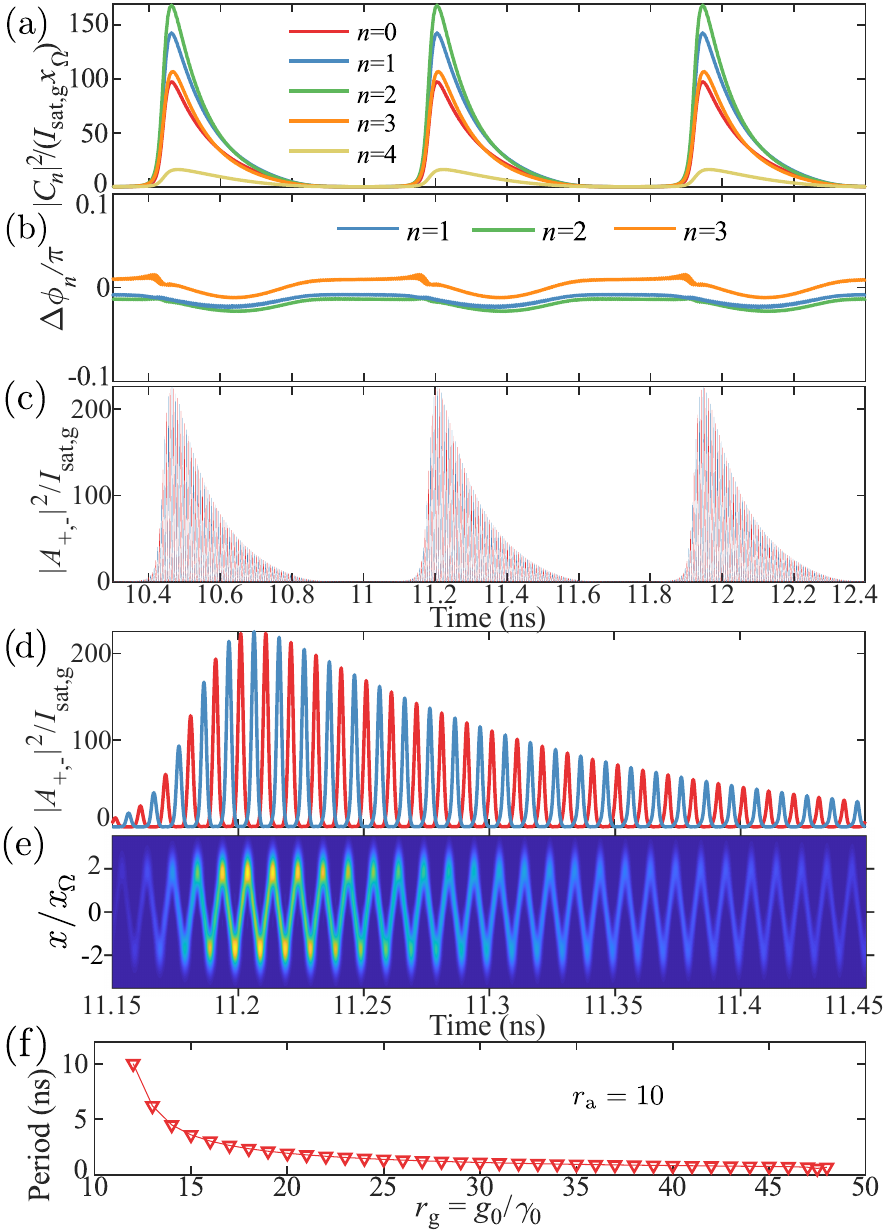}	
		\caption{Q-switched mode locked regime. Time evolutions of (a) the intensities of  modes $n=0..4$ (b) the relative phases $\Delta\phi_n = 2\phi_n-\phi_{n+1}-\phi_{n-1}$ between these modes and (c) the intensities  $|A_{+,-}|^2$ propagating in the $\pm x$ directions (plotted in red and blue lines, respectively) at cavity center $x=0$ in (c). (d) Zoom on one of the pulses of (c). (e) False color map of the  intensity $|A(x,t)|^2$ versus $t$ and $x$. The parameters are $r_\mathrm{g}=g_0/\gamma_0=45$, $r_\mathrm{a}=a_0/\gamma_0=10$, $w=5\,x_\Omega$, $\tau_g=1\,\mathrm{ns}$, $\tau_a=10\,\mathrm{ps}$. (f) Evolution of the Q-switching period with the gain.
		}
		\label{fig:Q_switched_mode_locking}
	\end{figure}
By increasing the unsaturated gain $g_0$, the recovery time of the gain between two Q-switch pulses is reduced, thus reducing the time between two such pulses. This then allows several modes to oscillate simultaneously  during one Q-switch pulse, and the phases of these modes can lock, leading to the Q-switched mode-locked operation regime labeled as (2) in Fig.\,\ref{fig:bifurcation_peaks}. One example of such a behavior is shown in greater details in Fig.\,\ref{fig:Q_switched_mode_locking} for $r_\mathrm{g}=g_0/\gamma_0=45$.  Figure\,\ref{fig:Q_switched_mode_locking}(a) reproduces the time evolutions of the intensities $|C_n|^2$ of the 5 lowest order modes. It shows that the delay between two Q-switch pulses is now reduced to less than 1\,ns, and that the 5 lowest order modes oscillate simultaneously during each pulse. Moreover, the relative phases $\Delta\phi_n = 2\phi_n-\phi_{n+1}-\phi_{n-1}$ between the modes with $n=1,2,3$ are locked to values close to 0, as evidenced in Fig.\,\ref{fig:Q_switched_mode_locking}(b).

The resulting intensities $|A_{+,-}|^2=|A_{\pm}(0,t)|^2$ for the part  of the fields at the cavity center $x=0$ propagating to the $+x$  (red line) and $-x$ (blue line)  directions are plotted versus in Fig.\,\ref{fig:Q_switched_mode_locking}(c). 
The shape of the Q-switch pulses is again asymmetric, like in Fig.\,\ref{fig:single_mode_q_switching}, but one can see that a every Q-switch pulse contains many much shorter pulses, which are formed by mode-locking. The zoom in Fig.\,\ref{fig:Q_switched_mode_locking}(d) shows actually that, as a result of the phase locking of the five HG modes, a single pulse with a duration of the order of 2~ps is bouncing back and forth inside the cavity during every Q-switch pulse. This is also clearly visible in the intensity color map of Fig.\,\ref{fig:Q_switched_mode_locking}(e). Moreover, Fig.\,\ref{fig:Q_switched_mode_locking}(f) shows that the period between two Q-switch pulses decreases with the laser excitation rate, as expected in standard Q-switched lasers.
	
	
\subsection{Continuous-wave mode locking}
	\begin{figure*}[htb]
		\centering
		\includegraphics[width=1\textwidth]{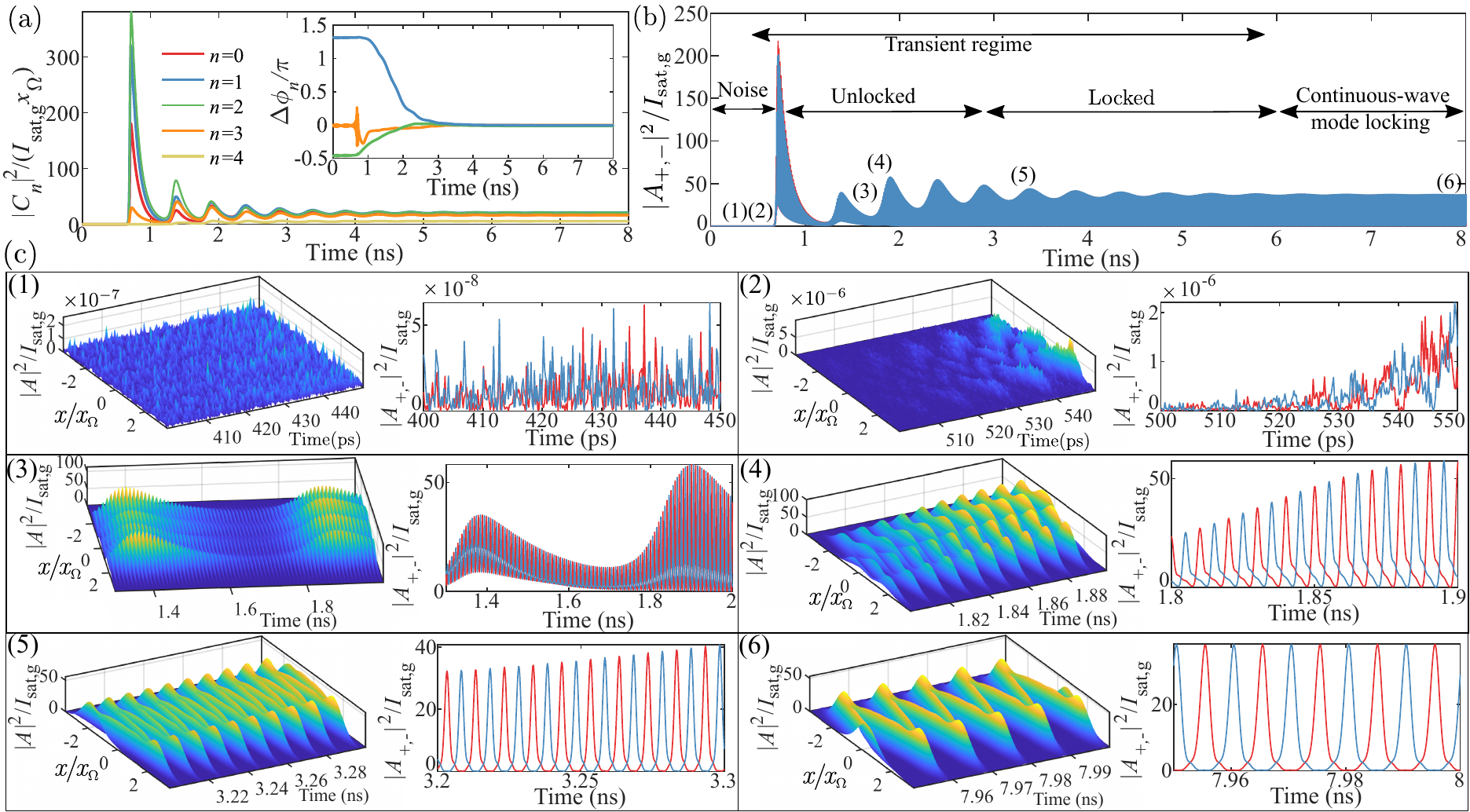}		
		\caption{Laser behavior during buildup of cw mode locking. (a) time evolution of the modes intensity and phase differences (inset).  (b) Time evolution of the intensity $|A_{+,-}|^2$ at  cavity center $x=0$. (c) Snapshots at 6 successive instants labeled as numbers in (b). 3D plot of the intracavity intensity $|A(x,t)|^2$  versus $x$ and $t$ and time evolution of the counterpropagating intensities $|A_{+,-}|^2$ at  cavity center $x=0$. Parameter values are $r_\mathrm{g}=g_0/\gamma_0=49$, $r_\mathrm{a}=a_0/\gamma_0=10$, $w=5\,x_\Omega$, $\tau_g=1\,\mathrm{ns}$, $\tau_a=10\,\mathrm{ps}$.}
		\label{fig:Continuous_wave_mode_locking_initial}
	\end{figure*}
	
By further increasing the pumping rate $g_0$, cw mode-locking can be observed. One example with $r_\mathrm{g}=g_0/\gamma_0=49$ is shown in Fig.\, \ref{fig:Continuous_wave_mode_locking_initial}. This figure shows the whole laser time evolution, from the noisy initial conditions to steady-state. Figure\, \ref{fig:Continuous_wave_mode_locking_initial}(a) shows the evolution of the mode intensities and, in inset, the evolution of the phase difference of the modes, while Fig.\, \ref{fig:Continuous_wave_mode_locking_initial}(b) shows the corresponding evolution  of the intensities  $|A_{+,-}|^2=|A_{+,-}(0,t)|^2$ at $x=0$ propagating in both directions. The small figures labeled (1-6) in Fig.\, \ref{fig:Continuous_wave_mode_locking_initial}(c) give snapshots of the laser behavior at several moments during laser buildup. 

 The whole transient process can be divided into several steps: 
 
Noise regime: the initial field is a random noise with very low intensities. The gain and absorber are activated at time of $t=0\,\rm{ns}$. The gain thus starts to  amplify the field after $t=450\,\rm{ps}$, as shown in (1). The gain goes on increasing and leads to significant amplification at around $t=500\,\rm{ps}$, as can be seen in (2). 
	
Transient regime: from $t=700\,\rm{ps}$ to $t\simeq6\,\rm{ns}$, the laser emits a series of spikes followed by relaxation oscillations. All first five modes oscillate simultaneously. The phase differences between the  modes are locked  around $t=3\,\rm{ns}$, as can be seen from the inset in Fig.\,\ref{fig:Continuous_wave_mode_locking_initial}(a) or by comparing the snapshots labeled (4) and (5) in Fig.\,\ref{fig:Continuous_wave_mode_locking_initial}(c). One can see that the pulse is not yet present in (4) while it is clearly there in (5). Furthermore, we notice that mode-locking occurs at $t\simeq3\,\rm{ns}$, much before the intensities of the modes reach their steady-state values ($t\simeq6\,\rm{ns}$). For example, the snapshot labeled (5) in Fig.\,\ref{fig:Continuous_wave_mode_locking_initial}(c) shows that the pulsed mode-locked regime is already well established although the laser power is still increasing versus time.
The duration of this transient damped oscillatory regime  depends on the pump rate $g_0$ and the absorption $a_0$. A stronger unsaturated gain $g_0$ leads to a faster transient, while a strong absorption  $a_0$  reduces the damping, thus lengthening the duration of the transient regime. If the absorption rate $a_0$ becomes too strong, the transient regime does not damp anymore and one retrieves the  Q-switched mode-locked regime of the preceding subsection (see Fig.\,\ref{fig:Q_switched_mode_locking}).
	
Stable mode-locking: cw mode locking is formed, where all mode intensities reach time independent values, phase differences between the modes are equal to 0. A pulse with stable intensity oscillates inside the cavity, with a repetition rate equal to $10\,\mathrm{ps}$, as shown in the snapshot labeled (6) in Fig.\, \ref{fig:Continuous_wave_mode_locking_initial}(c).
	
The cw mode locking is obtained thanks to an equilibrium between the effects of saturable gain and saturable absorber. As we have seen in Section \ref{Competition}, non-instantaneous gain saturation favors multimodes operation, which is a very important first step towards self-starting mode locking. Without any saturable absorber, mode  locking can also occur \cite{Rosales2012} through the nonlinearity of the saturable gain. But such a phase-locked operation does not always mean pulsed operation, for which one needs the relative phase difference $\Delta \phi$ between adjacent modes to be close to 0. This pulsed operation is strongly favored by the introduction of the saturable absorber. But if saturable absorption is too strong, the damping of relaxation oscillations doesn't occur, leading to instability of mode locking, i. e. appearance of Q-switched mode-locking. 

Finally, let us mention that the buildup of a mode-locked soliton in the harmonic cavity has some similarities with the buildup of a soliton in mode-locked fiber lasers, which can also be divided as several step: noise, beatnote, Q-switched beatnote, mode locking \cite{Liu2019, Liu2019b}.
	
	\section{Discussion}\label{Stability}
	
In this section, we discuss several features related to the mode-locked oscillation regime of the harmonic photonics cavity nanolaser, namely i) the role of the Henry factor in the gain and absorber media, ii) the possibility to spatially separate the gain and the absorber, and iii) the peculiarities of these lasers with respect to ordinary semiconductor lasers.

	\subsection{Role of Henry's factor}
	
The Henry factor quantifies the coupling between the variations of the real and imaginary parts of the active medium, which are respectively linked to the phase and amplitude variations in the laser.  In  semiconductor lasers, it plays an important role due to the particular dependence of the refractive index on the carrier density. If it is too large, the Henry factor can impede mode locking \cite{Vladimirov2005}. 

The robustness of mode-locking to the Henry factor in the harmonic photonic cavity laser is  investigated here by looking at the evolution of the spectrum of the laser in steady-state regime when the Henry factor increases. 
	
To this aim, the laser field spectrum $S(f)$ is calculated by the following expression:
	\begin{equation}
	S(f) =\int^{+\infty}_{-\infty}{\left|\int^{+\infty}_{-\infty}{A(x,t)e^{\mathrm{i}2\pi f t}dt}\right|^2dx}.
	\end{equation}
The mode frequencies $f_n$ can be obtained by extracting the peaks of the spectrum $S(f)$.
	
We simulate the behavior of the laser with $g_0/\gamma_0=70$, $a_0/\gamma_0=5$, $w=5\,x_\Omega$, $\tau_g=1\,\mathrm{ns}$, $\tau_a=10\,\mathrm{ps}$, and for varying values of the Henry factors $\alpha_{\rm g}$ and $\alpha_{\rm a}$ of the gain and absorber media that we suppose to be equal ($\alpha_{\rm g}=\alpha_{\rm a}$) for the sake of simplicity.

	\begin{figure}[htb]
		\centering
		\includegraphics[width=\columnwidth]{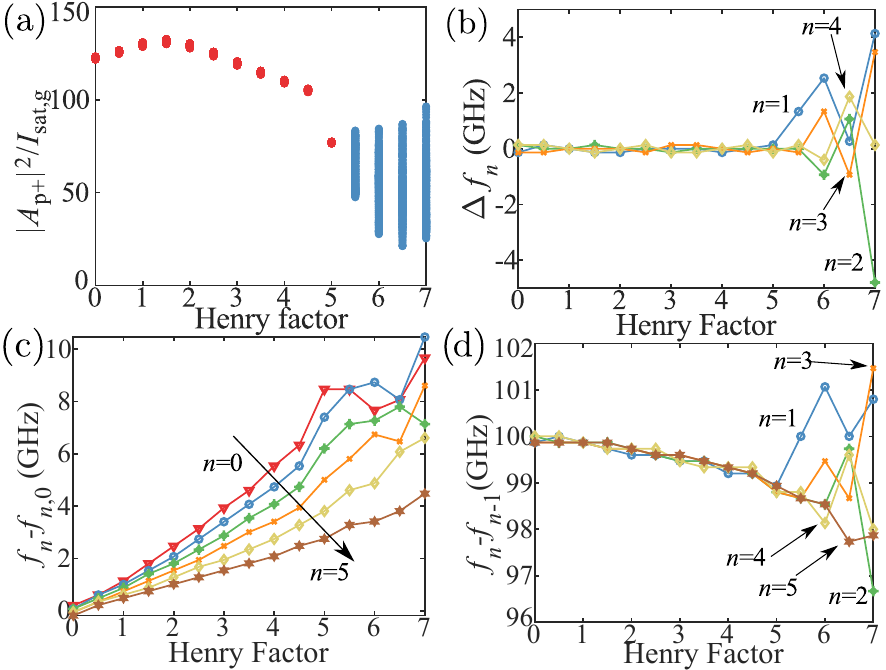}	
		\caption{Influence of the Henry factor on the laser steady-state behavior. (a) Intensity peaks for the field propagating in the $+x$ direction at cavity center $x=0$ as a function of the Henry factor $\alpha_{\rm g}=\alpha_{\rm a}$. (b) Mode Frequency difference $\Delta f_n = 2f_n-f_{n+1}-f_{n-1}$. (c) Mode frequency shifts from the empy cavity frequencies  $f_{n,0}=(n+1/2)\times 100\,\mathrm{GHz}$. (d)Mode separation $f_n-f_{n-1}$. The parameters are: $g_0/\gamma_0=70$, $a_0/\gamma_0=5$, $w=5\,x_\Omega$, $\tau_g=1\,\mathrm{ns}$, $\tau_a=10\,\mathrm{ps}$.}
		\label{fig:locking_henry_factor}
	\end{figure}
Figure\,\ref{fig:locking_henry_factor} shows the evolution of the steady-state laser behavior as a function of $\alpha_{\rm g}=\alpha_{\rm a}$. Figure\,\ref{fig:locking_henry_factor}(a) reproduces the peak  intensity $|A_{\rm p+}|^2$ of the field propagating in the $+x$ direction detected within a $10\,\rm{ns}$ time interval. It shows that the laser remains mode-locked as long as $\alpha_{\rm g}=\alpha_{\rm a}\leqslant5$. For larger values, the phases of the modes unlock, leading to the multiple blue dots. This behavior is confirmed by the evolution of the frequency differences $\Delta f_n = 2f_n-f_{n+1}-f_{n-1}$ between the modes reproduced in Fig.\,\ref{fig:locking_henry_factor}(b). Mode-locking corresponds to $\Delta f_n$ very close to 0. The corresponding evolution of the difference $f_n-f_{n,0}$ between the laser mode frequencies and the empty cavity frequencies $f_{n,0}=(n+1/2)\times 100\,\mathrm{GHz}$ shows that the Henry factor shifts the comb frequencies towards high frequencies (see Fig.\,\ref{fig:locking_henry_factor}(c)). This corresponds to a reduction of the mode separation $f_n-f_{n-1}$ of the comb, as can be seen in Fig.\,\ref{fig:locking_henry_factor}(d).

	\subsection{Asymmetric scheme for the gain and the absorber}\label{Asymmetric}
	
	\begin{figure}[htb]
		\centering
		\includegraphics[width=1\columnwidth]{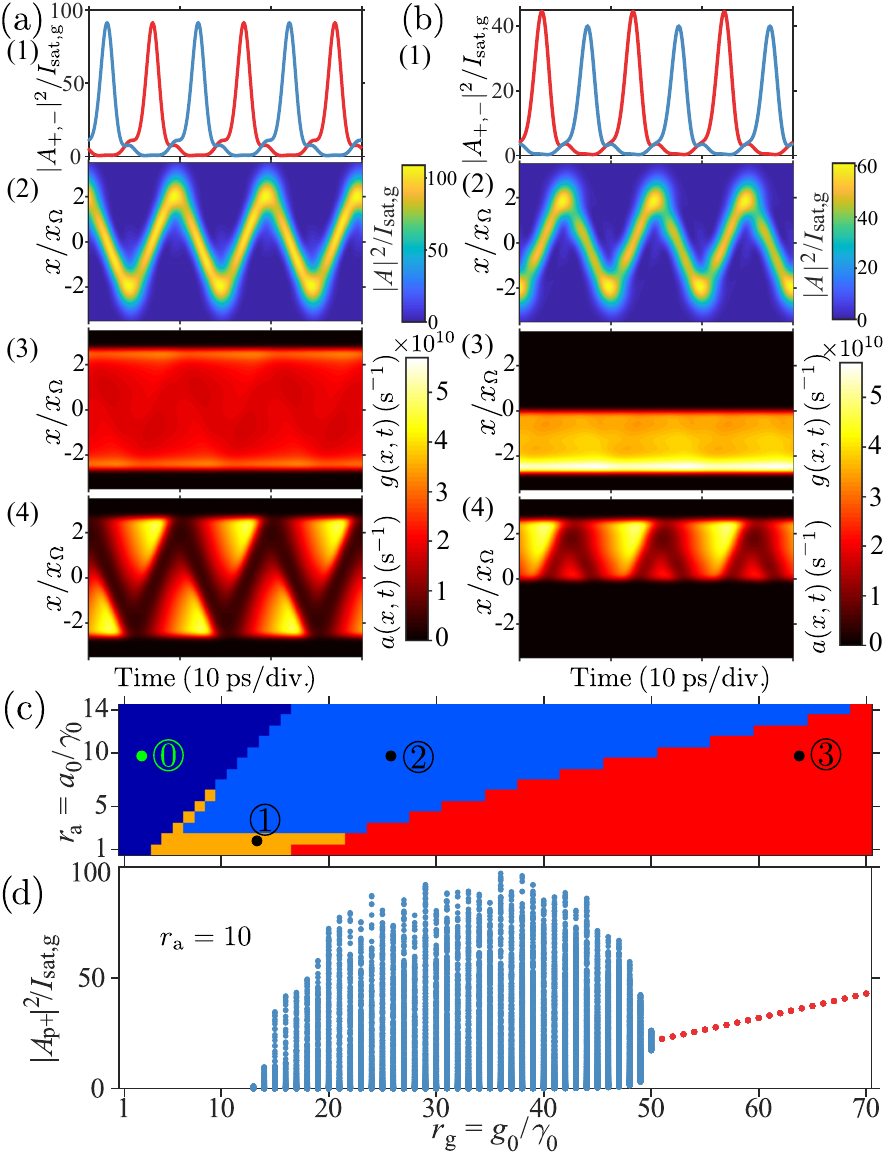}	
		\caption{
			(a,b) Steady-state mode-locked laser behavior for (a) superimposed and (b) separated gain and absorber regions. All other parameters are the same, with $r_{\mathrm{g}}=70$ and $r_{\mathrm{a}}=10$. (1) Time evolutions of the intensities $|A_{+,-}|^2$ of the fields traveling in the $+x$ (red line) and $-x$ (blue line) directions at cavity center ($x=0$); (2-4) False color maps of (2) intensity $|A(x,t)|^2$, (3) gain $g(x,t)$ (4) absorption $a(x,t)$. (c) Phase diagram representing the different steady-state regimes for separated gain and absorber regions as a function of the unsaturated gain and absorption: \textcircled{0} below threshold; \textcircled{1} multimode operation; \textcircled{2} Q-switching; \textcircled{3} cw mode locking. 
			(d) Corresponding bifurcation diagram:  peak values $|A_{\rm p+}|^2$ of the field intensity propagating in $+x$ direction at $x=0$ in  steady-state as a function of $r_{\rm{g}}$. The absorption is fixed at $r_\mathrm{a}=10$.
			Other parameter values are $w=5\,x_\Omega$, $\tau_g=1\,\mathrm{ns}$, $\tau_a=10\,\mathrm{ps}$.}
		\label{fig:separated_scheme}
	\end{figure}
Implementing a nanolaser based on a harmonic photonic resonator in which the gain and the saturable absorber are located at the same place is not always easy to do. We thus wonder in this subsection whether a laser in which the gain and the absorber are spatially separated may exhibit similar properties, in particular when it comes to the mode-locked regime, as in the scheme where gain and absorber overlap.

With the same parameters as before, we now consider the situation where the gain and the absorber do no longer overlap. The gain region is located in the region  $-2.5\,x_\Omega\leq x\leq 0$ and the absorber in the region  $0\leq x\leq 2.5\,x_\Omega$.
%
%

Figure \ref{fig:separated_scheme} shows a comparison of the laser steady-state mode-locked operation between the cases where the gain and absorber regions overlap (Fig.\,\ref{fig:separated_scheme}(a)) and where they are separated (Fig.\,\ref{fig:separated_scheme}(b)). All other parameters are the same, and in particular $r_{\rm g}=70$ and $r_{\rm a}=10$. In both cases, the panels labeled (1) show the time evolution of the intensities $|A_{+,-}|^2$ of the fields traveling in the $+x$ (red line) and $-x$ (blue line) directions at cavity center ($x=0$). Those labeled (2), (3), and (4) are false color maps of the intensity $|A(x,t)|^2$, the gain  $g(x,t)$, and the absorption $a(x,t)$, respectively.

One can see that the behaviors of the laser are similar for the two cases. The only significant difference is the difference between the peak values of $|A_{+}|^2$ and $|A_{-}|^2$ in case (b), and the reduction of the laser power due to the reduction of the size of the gain region. 

Another interesting feature can be seen by comparing panels (3) with panels (4): since the absorber is much faster that the gain, the ``tracks'' of the pulse can be seen in the former one but only hardly in the latter one.

In the case where the gain absorber regions are separated, the phase diagram obtained by tuning the gain $g_0$ and absorption $a_0$ is shown in Fig.\,\ref{fig:separated_scheme}(c). It is very similar to the former one (see Fig.\,\ref{fig:bifurcation_peaks}) and exhibits different behaviors, namely \textcircled{0} no lasing , \textcircled{1} multimode beating operation, \textcircled{2} Q-switching, and \textcircled{3} cw mode locking. 	

Region \textcircled{1}, which corresponds to cw multimode operation, happens when $r_a$ and $r_g$ are small, because in this case the saturable absorption and the intracavity power are not sufficient to sustain self-pulsing operation. Although Q-switching is observed in region \textcircled{2}, we notice that Q-switched mode-locked operation does not occur due to asymmetry between gain and absorption. 

Compared with the case of Fig.\,\ref{fig:bifurcation_peaks} where the gain and absorber overlap, we also notice that the transition line between regions \textcircled{2} and \textcircled{3} is shifted a little bit to higher gains. This is clear also on the bifurcation diagram shown in Fig.\,\ref{fig:separated_scheme}(d) with the fixed absorption coefficient $r_{\rm a}=10$. Compared with Fig. \ref{fig:bifurcation_peaks}, the peak values of the mode-locked pulse are about half of the overlapping scheme, and the transition point shifts from $r_g=48$ to $r_g=51$.

\subsection{Comparison with conventional semiconductor lasers}
	
There exist significant differences between the harmonic photonic cavity laser and the conventional Fabry-Perot semiconductor laser. The first one comes from the spatially inhomogeneous energy distribution of HG modes. A Fabry-Perot cavity sustains standing modes with equally spaced frequencies, the same length as the cavity, and their intensity is, apart from the nodes and antinodes, homogeneously distributed inside the resonator. 
	
On the contrary, the harmonic photonic cavity sustains HG modes whose 
spatial extension ranges with the square root of the mode order as shown in Fig.\,\ref{fig:mode_wide_and_gain_width}(a). 
The number of excited modes also depends on the width of the gain area due to the spatial inhomogeneity of the modes. Therefore, the length of the laser is mainly determined by the length of the active medium and does not affect the FSR of the cavity. This second difference is very helpful to reduce the size of the mode-locked laser while keeping a fixed value of the repetition rate.  

On the contrary, in a Fabry-Perot cavity, the FSR depends on the cavity length, and the orders of the excited modes are related to the gain spectrum. For example, a laser with FP cavity length equal to $430\,\rm\mu m$, center wavelength of $1.55\,\rm\mu m$, and refractive index equal to $3.5$, can sustain oscillation  of modes of order surrounding $1940$.

Taking the same value for the FSR (100\,GHz), we can illustrate the compactness of the concept of harmonic photonic cavity laser by considering that only very low order modes oscillate and are phase locked. This makes it feasible to reduce the scale of such a mode-locked semiconductor laser from sub-millimeter to few micrometers. For example, the size of the cavity in the preceding examples is equal to $5\,x_\Omega=42\,\mu m$.
	
Despite the strong differences between the modes of the two types of lasers,  we  find some similarities in their dynamic regimes, such as Q-switching, Q-switched mode locking, and cw mode locking  \cite{Javaloyes2010a,Vladimirov2005,Viktorov2007}. The reason behind these similarities is that mode locking is induced by the same physical mechanism in the two types of systems. 

To summarize this discussion, we believe that the interest of harmonic cavity nanolasers is that they are complementary to usual Fabry-Perot lasers, in the sense discussed above. Indeed, although different oscillation regimes, such as, e.g., passive mode-locking can be achieved in those nanolasers for the same reasons as in Fabry-Perot lasers (equally spaced modes and saturable absorption), the scaling of the pulse parameters (duration and repetition rate) with the cavity parameters are completely different in the two types of cavities. 

	\section{Conclusions}
We have derived and analyzed the possible dynamical behaviors of a nanolaser exhibiting Hermite-Gaussian modes created by a harmonic photonic cavity. Its special properties make it a promising candidate for the realization of mode-locked nanolasers. The FSR of the laser, and thus the repetition rate of the mode-locked pulses, is independent of the cavity length, but governed by the curvature of the photonic potential, so that the laser size mainly depends on the length of the active medium, giving this type of laser a strong advantage in terms of miniaturization. 
In addition, the fact that the Hermite-Gaussian modes that are locked are the lowest-order ones, i. e. the smallest ones, is very helpful to reduce the size of the mode-locked laser from  sub-millimeter to  micrometer range for a 100\,GHz repetition rate.

A  model based on the GPE with dissipative terms has been used to describe the effect of the harmonic cavity and of the active media, taking into account the spatial distribution of the gain, absorber, and HG modes.

To understand the saturation properties of the spatially inhomogeneous HG modes, we have compared the saturation matrices of the harmonic cavity nanolaser and the FP cavity laser, thus revealing that  cross-saturation of HG modes is predominant on  adjacent modes only. This has allowed us to understand the peculiarities of mode competition in such lasers. The steady-state behavior under non-instantaneous gain response depends on the gain length that limits the number of HG modes that can be excited and the mode intensity distribution. Strong differences have been observed with respect to the case of instantaneous gain saturation in which multi-stability has been observed. The transition from one regime to the other is related to the respective values of the  gain lifetime and the laser repetition time $2\pi/\Omega$.

Finally, we have isolated the different possible dynamical behaviors of the ML nanolaser obtained  by varying the gain and the absorption. These different regimes, including the Q-switching, Q-switched mode locking, and cw mode locking, were fully described illustrating the rich physics of this nonlinear system. In addition, the influence of the Henry factor on the mode locking has been discussed. Moreover, similar dynamical behaviors using spatially separated gain and absorber sections inside the cavity have been observed, which can simplify  practical implementations. Finally, a general comparison between harmonic cavity lasers and the traditional FP cavity lasers has been given.

\section*{Acknowledgments}
Work supported by the Direction G\'en\'erale de l'Armement and the Agence Nationale de la Recherche (LASAGNE, ANR-16-ASTR-0010-03), the ``Investissements d'Avenir'' program (CONDOR, ANR- 10-LABX-0035), European Union's Horizon 2020 program (Fun-COMP, Grant Agreement No. 780848) and performed in the framework of the joint research lab between TRT and LuMIn.

\end{document}